\documentclass[12pt]{article}
\usepackage{amsmath,amssymb}
\usepackage{graphicx}  
\usepackage{subfigure}
\usepackage{comment}
\usepackage{color}
\definecolor{darkblue}{rgb}{0.15,0.35,0.55}
\definecolor{reddish}{rgb}{0.65, 0.2, 0.2}
\definecolor{phthaloblue}{rgb}{0.0, 0.06, 0.54}
\definecolor{bluscuro}{rgb}{0.15, 0.2, .85}
\definecolor{rossos}{cmyk}{0,1,1,0.55}
\definecolor{bluchiaro}{cmyk}{1,.3,0.,0.1}
\usepackage[bookmarks,linktocpage, colorlinks=true, plainpages = false, citecolor = darkblue,  linkcolor=darkblue, urlcolor = darkblue, filecolor =darkblue]{hyperref} 
\usepackage{lipsum}
\usepackage{braket}

\begin{document}
\setcounter{footnote}{0}
\begin{titlepage}
\begin{center}
\hfill KEK-TH-1928\\
\vskip .75in

{\huge \bf
Gravitational relaxation of\\[0.7cm]
electroweak hierarchy problem}
\vskip .5in

{\large
  Hiroki Matsui$^{\rm (a,b)}$, Yoshio Matsumoto$^{\rm (b)}$, }
\vskip 0.25in

$^{\rm (a)}${\em 
KEK Theory Center, IPNS, KEK, Tsukuba, Ibaraki 305-0801, Japan}

\vskip 0.1in
$^{\rm (b)}${\em 
The Graduate University of Advanced Studies (Sokendai),Tsukuba, Ibaraki 305-0801, Japan}

\end{center}

\vskip .5in

\begin{abstract}
In the present paper, we discuss gravitational relaxation models 
for the electroweak hierarchy problem.
We show that modified gravity  
can naturally relax the electroweak hierarchy problem where 
conformal transformation
provides a crucial rule about what modified gravity theories are favored to relax
the electroweak hierarchy.
The conformal transformation connects different gravitational theories 
and rescaling the metric changes the dimensional parameters 
like the Higgs mass or the cosmological constant in different frames drastically.
When the electroweak scale is naturally realized by dynamical and running behavior of 
dilatonic scalar field or scaling parameter, the modified gravity theories 
can relax the electroweak hierarchy problem.
We discuss the theoretical and phenomenological validity of 
the gravitational relaxation models.
\end{abstract}

\end{titlepage}
\allowdisplaybreaks[1]

\tableofcontents

\section{Introduction}
The electroweak hierarchy problem has been recognized as the most notorious difficulty for 
the high-energy physics in the past decades and 
often rephrased as the naturalness problem~\cite{Wilson:1973jj,'tHooft:1979bh,Dine:2015xga,Giudice:2013nak}.
The naturalness that the low-energy effective field 
theory should not be extremely sensitive to the high-energy theory is a theoretical and reasonable presumption.
Actually, dimensional parameters like scalar masses are as large as the ultraviolet (UV) cut-off scale 
without involving any special fine-tuning of the parameters or any symmetry.
For the Standard Model (SM) case, the Higgs boson mass grows up to the UV cut-off scale ${ M }_{\rm   UV }$
by the quadratically divergent quantum corrections 
\footnote{
On the other hand, the cosmological constant problem is more serious 
from the viewpoint of the naturalness or hierarchy problem.
The quantum radiative corrections to the vacuum energy density ${ \rho  }_{ vacuum }$
which is dubbed \textit{zero-point vacuum energy} enlarges up to 
the cut-off scale ${ M }_{\rm  UV }$ as follows:
\begin{align*}
&\delta{ \rho  }_{ vacuum } =\frac{1}{2}\int^{{ M  }_{\rm   UV } } { \frac { { d }^{ 3 }k }{ { \left( 2\pi  \right)  }^{ 3 } }
 \sqrt { { k }^{ 2 }+{ m }^{ 2 } }  } =\frac { { M  }^{4}_{\rm  UV }}{ 16{ \pi  }^{ 2 } } +\frac { { { m }^{ 2 }{ M  }^{2}_{ \rm  UV } } }{ 16{ \pi  }^{ 2 } } 
 +\frac { { m }^{ 4 } }{ 64{ \pi  }^{ 2 } } \log { \left( \frac { { m }^{ 2 } }{ { M  }^{2}_{ \rm  UV } }  \right)  } +\cdots ,
\end{align*}
which is much larger than the dark energy 
$2.4\times 10^{-3}\  {\rm eV}$ in the current Universe.}
\begin{equation}
\delta M^{2}_{H}\simeq \frac{\alpha}{\left(4\pi\right)^{2}}{ M }^{2}_{\rm  UV },
\end{equation}
where $\delta M^{2}_{H}$ should not be much larger than the observed Higgs boson mass $M_{H}^{obs}=125.09\ {\rm GeV}$
\cite{Aad:2015zhl,Aad:2013wqa,Chatrchyan:2013mxa}.
The majority of theoretical efforts to solve the Higgs naturalness 
or electroweak hierarchy problem assume a TeV-scale new physics and many models 
has been proposed,
e.g. supersymmetry, extra-dimensions and compositeness.
However, these prominent proposals has been suffered 
from the observed Higgs boson mass $M_{H}^{obs}=125.09\ {\rm GeV}$ and 
the current experimental constraints on new physics.

Recently,  theoretically different approaches to the electroweak hierarchy problem 
have been explored in Ref.\cite{Graham:2015cka,2015PhRvL.115y1803E,
Hardy:2015laa,Patil:2015oxa,Batell:2015fma,DiChiara:2015euo,Matsedonskyi:2015xta,Evans:2016htp,Hook:2016mqo}
based on the cosmological relaxation model of Ref.\cite{Abbott:1984qf} 
explaining dynamically the smallness of the cosmological constant.
This relaxation mechanism is based on the cosmological evolution of the Higgs field 
and the axion-like field in the inflationary Universe,
and can lead to the naturally small electroweak scale against the cut-off scale.

In this paper we discuss other relaxation scenarios to the electroweak hierarchy problem 
by involving the gravitational modification which has been proposed and discussed
in Ref.\cite{Polyakov:1982ug,
Polyakov:2000fk,Jackiw:2005yc,Demir:2012nd,Lin:2014mua,Kobakhidze:2015jya,Kaloper:2013zca,
Kaloper:2014dqa,Kaloper:2014fca,Kaloper:2015jra,Kaloper:2016yfa,Kaloper:2016jsd}.
Especially, we demonstrate that these gravitational relaxation models 
can be embedded in the framework of modified gravity theory.
By using the conformal transformation or rescaling the metric
we clearly show that the electroweak hierarchy can be relaxed from 
the gravity sector. 
The conformal transformation connects different gravity theories 
and rescaling the metric drastically changes dimensional parameters 
like the Higgs boson mass or cosmological constant in the different frame.
We consider several gravitational relaxation models, and 
discuss the theoretical validity and phenomenological constraints.

\medskip
The layout of this paper is the following: In Section~\ref{sec:relaxation} 
we introduce the basic formulation for the gravitational relaxation scenarios and
discuss why the extended gravity theory relax the electroweak hierarchy problem
using the conformal transformation.
In Subsection~\ref{sec:sequestering} we apply the vacuum energy sequestering 
for the electroweak hierarchy problem
as an example of the gravitational modification.
In Section~\ref{sec:running-relaxation} we consider the running gravitational relaxation scenarios
where the quantum equivalence between
in the Einstein and Jordan frames is crucial. 
Finally, in Section~\ref{sec:conclusion} we summarize the conclusion of our work.

\section{Gravitational relaxation}
\label{sec:relaxation}
In this section, we introduce the gravitational relaxation scenario for the electroweak hierarchy problem.
We consider modified gravity theory
including the Higgs field $\Phi$ and the dilatonic scalar field $\chi$ non-minimally coupled to the gravity. 
The classic action is written by
\begin{equation}
S \supset  S_{\rm gravity}+S_{\rm Higgs}\label{eq:sdkjsldk}.
\end{equation}
The action for the gravity sector including the dilatonic scalar field $\chi$ is given by~\cite{Kaiser:2010ps}
\begin{eqnarray}
\begin{split}
S_{\rm gravity} = &\int { { d }^{ 4 }x\sqrt { -g } }  \biggr( F\left(\chi\right)R  
-\frac {G\left(\chi\right) }{ 2 } { g }^{ \mu \nu  }
{ \nabla  }_{ \mu  }{ \chi}{ \nabla  }_{ \nu  }\chi -V\left(\chi\right)  \biggr) ,
\end{split}\label{eq:sdedg}
\end{eqnarray}
The action for the Higgs sector is given by
\begin{equation}
S_{\rm Higgs}=-\int { { d }^{ 4 }x\sqrt { -g }  \left( \frac { 1 }{ 2 } { g }^{ \mu \nu  }
{ \nabla   }_{ \mu  }{ \Phi }^{ \dagger  } {\nabla  }_{ \nu  }\Phi  +V\left({ \Phi  }^{ \dagger  }\Phi  \right)  \right)   } ,
\end{equation}
where the (bare) Higgs potential can be written by
\begin{equation}
V\left({ \Phi  }^{ \dagger  }\Phi  \right) =\Lambda_{\rm b}+M^{2}_{\Phi }\left({ \Phi  }^{ \dagger  }\Phi  \right)
+\lambda \left({ \Phi  }^{ \dagger  }\Phi  \right)^{2},
\end{equation}
Now, we assume that the Higgs mass $M_{\Phi }$ and 
the cosmological constant $\Lambda_{\rm b}$ are the UV scale to be
$\Lambda^{1/4}_{\rm b} \simeq M_{\Phi }\simeq M_{\rm UV}$.
If there exists exact supersymmetry or conformal symmetry, 
such symmetries can force the cosmological constant or the Higgs mass parameter to be smaller than the cut-off scale. 
However, such symmetries are always broken in real world and the dimensional parameters grow in proportion to the breaking scale. 
Thus, we eventually encounter the hierarchy problem via the symmetry breaking scale, 
and that is the situation in the standard SUSY models. 
Generally, it is difficult to protect the dimensional parameters from both large classical and 
quantum corrections of the high-energy physics and that is the reason why the 
hierarchy problem is thought to be serious.

Now, we rescale the metric via the conformal transformations as follows~\cite{Kaiser:2010ps}:
\begin{eqnarray}
g_{\mu\nu}&\rightarrow&\overline { g }_{\mu\nu} =\Omega^{2}\left(\chi\right) g_{\mu\nu},\\
g^{\mu\nu}&\rightarrow& \overline { g }^{\mu\nu} =\Omega^{-2}\left(\chi\right) g^{\mu\nu},\\
\sqrt { -g  } &\rightarrow& \sqrt { -\overline{ g } }=\Omega^{4}\left(\chi\right) \sqrt { -g  }.
\end{eqnarray}
The scalar curvature is transformed as follows:
\begin{eqnarray} 
\overline { R } = \frac{1}{ \Omega^{2}\left(\chi\right)} \left[  R-\frac{6\Box \Omega\left(\chi\right) }{\Omega\left(\chi\right)}\right] ,
\end{eqnarray}
where $\Box$ denotes the covariant d'Alembertian operator and satisfies,
\begin{eqnarray}
\Box \Omega=g^{\mu\nu}{ \nabla }_{ \mu  }
{ \nabla  }_{ \nu  }{ \Omega}=\frac{1}{\sqrt { -g }}
{ \partial  }_{ \mu  }\left[\sqrt { -g }g^{\mu\nu} { \partial  }_{ \nu  }{ \Omega}  \right] .
\end{eqnarray}
Thus, the action for the gravity sector can be transformed as follows:
\begin{eqnarray}
\begin{split}
S_{\rm gravity} = &\int { { d }^{ 4 }x\sqrt { -\overline { g } } }  \biggr( \frac { M_{\rm pl}^{2}}{ 2 }\overline { R } -\frac { \overline {G}\left(\chi \right) }{ 2 } \overline{ g }^{\mu\nu}
{ \overline\nabla  }_{ \mu  }{\chi}{\overline \nabla  }_{ \nu  }\chi -\overline{V}\left(\chi \right)  \biggr),
\end{split}\label{eq:sdlsjedg}
\end{eqnarray}
where we write down the action in Einstein frame where the scalar curvature is not multiplied by the scalar field,
and $\Omega \left(\chi\right)$, ${F}\left(\chi \right)$, $\overline {V}\left(\chi \right)$ and $\overline {G}\left(\chi \right)$ are given by
\begin{align}
&\Omega^{2}\left(\chi\right) =\frac{2 F\left(\chi\right)}{M_{\rm pl}^{2}} ,\quad
\overline {V}\left(\chi \right)=\frac{V\left(\chi\right) }{\Omega^{4}\left(\chi\right)}, \\
&\overline {G}\left(\chi \right)=\frac{G\left(\chi\right)}{\Omega^{2}\left(\chi\right)}
+\frac{6M_{\rm pl}^{2}}{\Omega^{2} \left(\chi\right)}\frac{ {\nabla  }_{ \mu  }
{\Omega}{ \nabla  }_{ \nu  }\Omega  }{ {\nabla  }_{ \mu  }{\chi}{ \nabla  }_{ \nu  }\chi  }+\cdots.
\end{align}
where $M_{\rm pl}=1/(8\pi G)^{1/2}=2.4\times10^{18}\ {\rm GeV}$ is
the reduced Planck mass and 
the related Newton's constant has tight constraints from the cosmological observations~\cite{Umezu:2005ee,Galli:2009pr}.

For the Higgs sector the action is given by
\begin{eqnarray}
\begin{split}
S_{{\rm Higgs}} = &-\int { { d }^{ 4 }x\sqrt { -g } }  \biggr( \frac { 1 }{ 2 } { g }^{ \mu \nu  }
{ \nabla   }_{ \mu  }{ \Phi }^{ \dagger  } {\nabla  }_{ \nu  }\Phi  
+\Lambda_{\rm b}+M^{2}_{\Phi }\left({ \Phi  }^{ \dagger  }\Phi  \right)+\lambda \left({ \Phi  }^{ \dagger  }\Phi  \right)^{2}\biggr) ,
\end{split}
\end{eqnarray}
The Higgs field $\Phi$ are transformed as 
\begin{eqnarray}
\Phi \rightarrow H =\Omega^{-1} \left(\chi\right)\Phi ,
\end{eqnarray}
The Higgs potential is transformed as follows:
\begin{equation}
V\left({H}^{ \dagger  }H  \right) =\frac{\Lambda_{\rm b}}{\Omega^{4} \left(\chi\right)}
+\frac{M^{2}_{\Phi }}{\Omega^{2} \left(\chi\right)}\left({H}^{ \dagger  }H \right)+\lambda \left({H}^{ \dagger  }H \right)^{2}.
\end{equation}

Note that the action in the SM is conformally invariant except for the Higgs potential.
On the other hand, the action of the gauge or fermion fields 
are only rescaled via the conformal transformations and the couplings are not changed.
As the mathematical manipulation, there are several metric frames,
e.g. Jordan frame (String frame) and Einstein frame.
However, we comment that there is no consensus about physical equivalence of these frames over the years~\cite{Faraoni:1999hp,
Capozziello:2010sc,Capozziello:2012nr,Deruelle:2010ht,Steinwachs:2013tr,Calmet:2012eq,
Domenech:2015qoa,Postma:2014vaa}
and one should not determine a unique physical frame in modified gravity theory.

The modified gravity theory like 
extra-dimensions or string theory is often written by Jordan frame.
The conformal transformation from Jordan frame to Einstein frame
suppress the dimensional parameters of the ordinary SM via the scaling parameter.
Our set-up is similar to the Randall-Sundrum model~\cite{Randall:1999ee}
where the large hierarchy is suppressed by the exponential warping factor ${ e }^{ -k{ r }_{ c }\phi  }$ which depends on an addition extra-dimension
\footnote{
The five-dimensional metric in the Randall-Sundrum model takes the form       
\begin{align*}
{ ds }^{ 2 }={ e }^{ -2k{ r }_{ c }\phi  }{ \eta  }_{ \mu \nu  }{ d }x^{ \mu  }{ d }x^{ \nu  }+{ r }_{ c }^{ 2 }d{ \phi  }^{ 2 } ,
\end{align*}
where ${ \eta  }_{ \mu \nu  }$ is the 4D Minkowski metric. }.
In this scenario, the four-dimensional components of the bulk metric ${ g }^{\mu\nu}$ and 
the four-dimensional physical metric $\overline{ g }^{\mu\nu}$ have the relation
${ g }^{\mu\nu}= { e }^{ -2k{ r }_{ c }\phi  }\overline{ g }^{\mu\nu}$
where $k$ is a Planck scale constant and $\phi$ is the extra-dimensional coordinate with the size ${ r }_{ c }$
\footnote{
In the Randall-Sundrum model, the four-dimensional action for the gravity sector is given by       
\begin{align*}
{ S }_{\rm gravity }\supset \int { { d }^{ 4 }x } \int { d\phi  }\ 2{ M }^{ 3 }{ r }_{ c }{ e }^{ -2k{ r }_{ c }\left| \phi  \right|  }\sqrt { -\overline { g }  }\ \overline { R },
\end{align*}
where $M$ is the five-dimensional Planck scale and $\overline { R }$ is constructed by the rescaling metric $\overline { g }_{ \mu \nu  } $.
Ther 4D Planck scale $M_{\rm pl}$ can be determined as follows:
 \begin{align*}
M_{\rm pl}^{2}= { M }^{ 3 }{ r }_{ c } \int { d\phi  }\ { e }^{ -2k{ r }_{ c }\left| \phi  \right|  }= \frac { { M }^{ 3 } }{ k } \left\{ 1-{ e }^{ -2k{ r }_{ c }\pi  } \right\}.
\end{align*}}.
By using the rescaling physical metric $\overline{ g }^{\mu\nu}$ 
instead of the bulk metric ${ g }^{\mu\nu}$,
this action can be written as,
\begin{eqnarray}
\begin{split}
S_{{\rm Higgs}} = &-\int { { d }^{ 4 }x\sqrt { -\overline{ g } } } { e }^{ -4k{ r }_{ c }\phi  } \biggr( \frac { { e }^{ 2k{ r }_{ c }\phi  } }{ 2 }\overline{ g }^{ \mu \nu  }
{ \overline\nabla  }_{ \mu  }{ \Phi}{\overline \nabla  }_{ \nu  } \Phi 
+\Lambda_{\rm b}+M^{2}_{\Phi }\left({ \Phi  }^{ \dagger  }\Phi  \right)+\lambda \left({ \Phi  }^{ \dagger  }\Phi  \right)^{2}\biggr) \\
= &-\int { { d }^{ 4 }x\sqrt { -\overline{ g } } } \biggr( \frac { 1 }{ 2 }\overline{ g }^{ \mu \nu  }
{ \overline\nabla  }_{ \mu  }{ H }{\overline \nabla  }_{ \nu  }H 
+\Lambda+M^{2}_{H }\left({ H  }^{ \dagger  }H \right)+\lambda \left({ H }^{ \dagger  }H \right)^{2}\biggr),
\end{split}
\end{eqnarray}
where $H$ is the transformed Higgs field satisfying the relation
$H ={ e }^{ k{ r }_{ c }\phi  }\Phi $. The Higgs mass parameter $M_{\Phi }$ in the fundamental higher-dimensional theory 
can be suppressed when measured with the rescaling physical metric $\overline{ g }^{\mu\nu}$
and become the order of the electroweak scale without difficulty as the following,
\begin{equation}
M_{H }= { e }^{ -k{ r }_{ c }\phi  }M_{\Phi }
\end{equation}

Let us discuss several gravitational relaxation models.
If we take $ \Omega^{2} \left(\chi\right)=M_{\rm pl}^{2}/\chi^{2}$~\cite{Wetterich:2002wm,Wetterich:2013wza} 
and ${F}\left(\chi \right)=M_{\rm pl}^{4}/2\chi^{2}$, 
the transformed Higgs potential is given by
\begin{equation}
V\left({H}^{ \dagger  }H  \right) =\frac{\Lambda_{\rm b}}{M_{\rm pl}^{4}}\chi^{4}+\frac{M^{2}_{\Phi }}{M_{\rm pl}^{2}}\chi^{2}
\left({H}^{ \dagger  }H \right)+\lambda \left({H}^{ \dagger  }H \right)^{2}.
\end{equation}
where the Higgs mass and cosmological constant 
are suppressed and screened off from the cut-off scale.
Next, we consider $ \Omega^{2} \left(\chi\right)=\chi^{2}/M_{\rm pl}^{2}$,
${F}\left(\chi \right)=\chi^{2}/2$,
the transformed Higgs potential can be written as
\begin{equation}
V\left({H}^{ \dagger  }H  \right) =\frac{M_{\rm pl}^{4}\Lambda}{\chi^{4}}+\frac{M_{\rm pl}^{2}M^{2}_{\Phi }}{\chi^{2}}
\left({H}^{ \dagger  }H \right)+\lambda \left({H}^{ \dagger  }H \right)^{2}.
\end{equation}
which correspond to the screen scenario for the cosmological constant 
by so-called cosmon field~\cite{Wetterich:1987fm,Wetterich:1994bg,Wetterich:2002wm,Wetterich:2013wza}.
When the classic scalar field or vacuum expectation value (VEV) become larger and larger
$\chi \gg 1 $,
the Higgs mass or the cosmological constant are sufficiently 
suppressed and asymptotically vanish.
Next we consider the specific dilaton model~\cite{Lin:2014mua} where the action can be written as follows:
\begin{eqnarray}
\begin{split}
S_{{\rm gravity}+{\rm Higgs}} = &\int { { d }^{ 4 }x\sqrt { -g } }  \biggr( \frac { M_{ \rm  UV}^{2}}{ 2}e^{2\chi/ \eta}R 
- \frac{1}{2}{ g }^{ \mu \nu  }{ \nabla  }_{ \mu  }{ \chi}{ \nabla  }_{ \nu  }\chi  \\ &
 -{ \lambda  }_{ \chi  }\left( { \chi  }^{ 2 }-{ v }_{ \chi  }^{ 2 } \right) -\frac { 1 }{ 2 } { g }^{ \mu \nu  }
{ \nabla   }_{ \mu  }{ \Phi }^{ \dagger  } {\nabla  }_{ \nu  }\Phi 
-\Lambda_{\rm b}-M^{2}_{\Phi }\left({ \Phi  }^{ \dagger  }\Phi  \right)-\lambda \left({ \Phi  }^{ \dagger  }\Phi  \right)^{2}\biggr) .
\end{split}
\end{eqnarray}
where we assume that $M$ is the order of the cut-off scale to be
$\Lambda^{1/4}_{\rm b} \simeq M_{\Phi }\simeq  M_{ \rm  UV}$.
Let us transform this action into Einstein frame by rescaling the metric,
\begin{equation}
\overline { g }_{\mu\nu} =\frac { { M }_{ \rm  UV}^{ 2 } }
{ M_{\rm pl}^{ 2 } } { e }^{ 2\chi /\eta  } g_{\mu\nu}
\simeq \frac { { M }_{ \rm  UV}^{ 2 } }{ M_{\rm pl}^{ 2 } } { e }^{ 2{ v }_{ \chi  } /\eta  } g_{\mu\nu}.
\end{equation}
The Higgs mass parameter $M_{\Phi }$ can be exponentially suppressed as the following
\begin{equation}
M_{H }\simeq \frac { { M }_{\rm pl} M_{\Phi }}{ M_{ \rm  UV} } { e }^{ -{ v }_{ \chi  } /\eta  }
\end{equation}
where we can have two approaches and interpretations.
If we regard ${ g }^{\mu\nu}$ as the physical metric, 
the Planck-mass scale emerges dynamically by the spontaneous symmetry breaking of the dilaton symmetry as
${ M }_{\rm pl}\simeq M_{ \rm  UV} { e }^{ { v }_{ \chi  } /\eta  }$.
We can solve the large hierarchy problem by 
assuming $\Lambda^{1/4}_{\rm b} \simeq M_{\Phi }
\simeq M_{ \rm  UV} \simeq  M_{\rm EW}$ where 
$M_{\rm EW}$ express the electroweak scale.
On the other hand, we can regard $\overline{ g }^{\mu\nu}$ as the physical metric
and set $M_{\rm EW }\simeq { M }_{\rm pl} { e }^{ -{ v }_{ \chi  } /\eta  }$.
This approach resembles the 
Randall-Sundrum model and the simplest possibility of gravitational relaxations
although the quantum gravity effects might appear 
above the electroweak scale~\cite{Lin:2014mua}.
In these models, the dimensional parameters like the Higgs boson mass, or even the cosmological constant
are screened off from the cut-off scale.
However, we can not solve both the electroweak hierarchy problem
and the cosmological constant problem at once.
After all we encounter the fine-tuning problem between the TeV scale and the dark energy 
although the physical cosmological constant to be $\Lambda^{1/4} \simeq \Omega^{-1}\left(\chi\right)\Lambda^{1/4}_{\rm b}
\simeq  M_{\rm EW}$ might be more or less relaxed.
Thus, we must require another relaxation mechanism for the cosmological constant.

\subsection{Sequestered electroweak hierarchy}
\label{sec:sequestering}
Hereafter, we will show that the vacuum energy sequestering scenario 
can also be effective against the electroweak hierarchy problem 
and discuss the relation between such a model 
and previous scenarios.
The vacuum energy sequestering scenario proposed in 
Ref.\cite{Kaloper:2013zca,Kaloper:2014dqa,Kaloper:2014fca,Kaloper:2015jra,
Kaloper:2016yfa,Kaloper:2016jsd} is the simple model
to solve the cosmological constant problem, and can relax the large discrepancy between 
the vacuum energy density from quantum corrections and the current observed value via the scaling parameter $\eta$.
This scenario assumes a minimal modification of general relativity 
to make all scales in the matter sector functionals of the 4-volume element of the Universe.
In the context of this scenario, 
the Universe should be finite in space-time and a transient stage with the present epoch of accelerated expansion
before the big crunch~\cite{Avelino:2014nqa}, but it has been shown that these models could be consistent with the cosmological observation
(there are similar models in the context of 
the unimodular gravity and more detailed discussions are given by 
Ref.\cite{Anderson:1971pn,Buchmuller:1988wx,Buchmuller:1988yn,Henneaux:1989zc,
Unruh:1988in,Nojiri:2016mlb,Nojiri:2016ygo,Nojiri:2016ppu,Nojiri:2016plt}).
The mechanism of this scenario is almost the same as the previously discussed one and 
the unknown scaling parameter $\eta$ can be regarded as the VEV of the 
dilatonic scalar field. 
The vacuum energy sequestering scenario is described by the following action
\cite{Kaloper:2013zca}
\begin{eqnarray}
\begin{split}
S_{{\rm gravity}+{\rm Higgs}}=&\int { { d }^{ 4 }x }\sqrt { -g } \left[ \frac { { M }_{ \rm pl }^{ 2 } }{ 2 } R 
-\Lambda_{\rm b}-{ \eta  }^{ 4 }\mathcal{L} 
\left( {  \eta    }^{ -2 }{ g }^{ \mu \nu  },\Phi  \right)  \right]  
+\sigma \left( \frac { \Lambda_{\rm b}  }{ \eta^{4} \mu^{4}  }  \right),
\end{split}
\end{eqnarray}
where $\Lambda_{\rm b}$ is the bare cosmological constant and $ \eta$ is the scaling parameter 
relaxing the large hierarchy. The function $\sigma (x)$ is  an adequate function to impose
the global constraints and $\mu$ is a parameter with the mass dimension.
The Lagrangian density $\mathcal{L} \left( {  \eta    }^{ -2 }{ g }^{ \mu \nu  },\Phi  \right)$ for the matter sector couples minimally to the rescaled metric 
$\overline { g }_{\mu\nu} = \eta^{2}{ g }_{\mu\nu}$ and includes the Higgs potential as
\begin{eqnarray}
\mathcal{L} \left( {  \eta    }^{ -2 }{ g }^{ \mu \nu  },\Phi  \right)&\supset&
V\left({ \Phi  }^{ \dagger  }\Phi  \right) \nonumber \\ 
&=&\Lambda_{\rm b}+M^{2}_{\Phi }\left({ \Phi  }^{ \dagger  }\Phi  \right)
+\lambda \left({ \Phi  }^{ \dagger  }\Phi  \right)^{2}.
\end{eqnarray}
The parameter $ \eta$ sets the hierarchy between the physical scale 
(the electroweak scale or the dark energy scale) and the UV cut-off scale $M_{\rm UV}$.
Thus, the Higgs mass parameter of the order of the UV cut-off scale are sufficiently suppressed as 
\begin{equation}
\frac{M_{H}}{M_{\rm  UV}}=\eta \frac{M_{\Phi}}{M_{\rm  UV}},
\end{equation}
where $M_{H}$ is the observed Higgs boson mass and $\eta \ll1$.
To show more accurately the heart of this mechanism, we rewrite this action for the Higgs sector by using the rescaled metric $\overline { g }_{\mu\nu} $ as
\begin{eqnarray}
\begin{split}
S_{{\rm gravity}+{\rm Higgs}} =&\int { { d }^{ 4 }x }\sqrt { -g } \left( \frac { { M }_{ \rm pl }^{ 2 } }{ 2 } R 
-\Lambda_{\rm b}  \right) \\ & 
-\int { { d }^{ 4 }x }\sqrt { -\overline{g} } \mathcal{L} 
\left( { \overline{g} }^{ \mu \nu  },\Phi  \right) +\sigma \left( \frac { \Lambda_{\rm b}  }{ \eta^{4} \mu^{4}  }  \right),
\end{split}
\end{eqnarray}
where
\begin{equation}
\sqrt { -\overline {g} } \mathcal{L}
\left( { \overline {g} }^{ \mu \nu  },\Phi  \right) = \sqrt { -g }{ \eta  }^{ 4 }\mathcal{L} 
\left( {  \eta    }^{ -2 }{ g }^{ \mu \nu  },\Phi  \right).
\end{equation}
The matter Lagrangian is written by the rescaling metric $\overline { g }_{\mu\nu} $ as 
\begin{eqnarray}
\begin{split}
\sqrt { -\overline {g} } \mathcal{L} \left( { \overline{g} }^{ \mu \nu  },\Phi  \right)=&\sqrt { -\overline{g} }
\biggr[ \frac { 1 }{ 2 } {\overline{g} }^{ \mu \nu  }{ \overline\nabla   }_{ \mu  }{ \Phi }^{ \dagger  } {\overline \nabla  }_{ \nu  }\Phi    
+\Lambda +M^{2}_{\Phi }\left({ \Phi  }^{ \dagger  }\Phi  \right)+\lambda \left({ \Phi  }^{ \dagger  }\Phi  \right)^{2}\biggr] .
\nonumber 
\end{split}
\end{eqnarray}
Thus, by using the conformal transformation, the matter Lagrangian can be written as
\begin{eqnarray}
\begin{split}
\sqrt { -\overline {g} } \mathcal{L} 
\left( { \overline {g} }^{ \mu \nu  },\Phi  \right) &=
\sqrt { -g }{ \eta  }^{ 4 }\mathcal{L} 
\left( {  \eta    }^{ -2 }{ g }^{ \mu \nu  },\Phi  \right) \\ &=\sqrt { -g }
\biggr[ \frac { 1 }{ 2 } { g }^{ \mu \nu  }{ \nabla   }_{ \mu  }{ H}^{ \dagger  } {\nabla  }_{ \nu  }H  
+\eta^{4}\Lambda +\eta^{2}M^{2}_{\Phi }\left({ H  }^{ \dagger  }H  \right)
+\lambda \left({ H }^{ \dagger  }H  \right)^{2}\biggr] ,\nonumber
\end{split}
\end{eqnarray}
where we assume $H=\eta \Phi$. 
Therefore, if the Higgs sector is sequestered from the gravitational sector via the scaling parameter $\eta \ll1$,
the large Higgs mass parameter can be sufficiently suppressed.
The unknown scaling parameter $\eta $ can be regarded as the dilatonic scalar field 
in the scenarios previously discussed.
Although the cosmological constraints on the modified gravity theory
and the equivalence between the Einstein and Jordan frames 
should be carefully considered, this mechanisms or scenarios would 
not significantly change. In this paper 
we have focused on the possibility relaxing the electroweak hierarchy
and left detail discussion on the cosmological constraints for a forthcoming work.

\section{Running gravitational relaxation}
\label{sec:running-relaxation}
In the previous section, we have discussed how modified gravity can 
alleviate the large Higgs mass and the cosmological constant by using the 
conformal transformation.
In this section, we discuss the running gravitational relaxation scenario
\footnote{
Polyakov proposed that the cosmological constant could be 
screened by the IR behavior of quantum gravity and 
the behavior can be translated by the
RG running of the auxiliary gravitational field~\cite{Polyakov:1982ug,Polyakov:2000fk}.
The electroweak hierarchy is also discussed by Ref.~\cite{Kobakhidze:2015jya}.}
in which we treat the dilatonic scalar field $\chi$ as the quantum field
and consider the renormalisation group (RG) running behaviors

For simplicity, we consider the action for the Higgs sector in Jordan frame as follows:
\begin{eqnarray}
\begin{split}
S_{{\rm gravity}+{\rm Higgs}} \supset &\int { { d }^{ 4 }x\sqrt { -g } }  \biggr( \frac { M_{\rm pl}^{4}}{ 2\chi^{2}}R-\frac { 1 }{ 2 } { g }^{ \mu \nu  }
{ \nabla   }_{ \mu  }{ \Phi }^{ \dagger  } {\nabla  }_{ \nu  }\Phi  
-\Lambda_{\rm b} - M^{2}_{\Phi }\left({ \Phi  }^{ \dagger  }\Phi  \right)-\lambda \left({ \Phi  }^{ \dagger  }\Phi  \right)^{2}\biggr) ,
\end{split}\label{eq:sdedsdg}
\end{eqnarray}
where we assume that the dilatonic scalar $\chi$ satisfy 
\begin{equation}\label{eq:sfggdg}
G\left(\chi \right)=6M_{\rm pl}^{2}/\chi^{2}\left(2-M_{\rm pl}^{2}/\chi^{2}\right)+\cdots, \ {V}\left(\chi \right)=0.
\end{equation}
This set-up is for simplifying our discussion and this action is
consistent with induced 
gravity theories~\cite{PhysRevLett.42.417,PhysRevLett.44.703,
Smolin:1979uz,Adler:1982ri,Dehnen:1989vg,Dehnen:1992rr,Spokoiny:1984bd,Accetta:1985du,Lucchin:1985ip,
Fakir:1990iu,Kaiser:1993bq,Kaiser:1994wj,CervantesCota:1994zf,CervantesCota:1995tz}

Let us perform the conformal transformation 
$ \Omega^{2} \left(\chi\right)=M_{\rm pl}^{2}/\chi^{2}$ and 
consider the action for the Higgs sector in Einstein frame 
\begin{eqnarray}
\begin{split}
S_{{\rm gravity}+{\rm Higgs}} \supset &\int { { d }^{ 4 }x\sqrt { -\overline{g} } }  \biggr( \frac { M_{\rm pl}^{2}}{ 2}\overline{R }
- 6\overline{ g }^{ \mu \nu  }{ \overline{\nabla}  }_{ \mu  }{ \chi}{ \overline{\nabla} }_{ \nu  }\chi  \\ &
-\frac { 1 }{ 2 } \overline{ g }^{ \mu \nu  }
{ \overline{\nabla} }_{ \mu  }{ H}^{ \dagger  } {\overline{\nabla} }_{ \nu  }H-\frac{\Lambda_{\rm b}}{M_{\rm pl }^{4}}\chi^{4} -\frac{M^{2}_{\Phi }}{M_{\rm pl }^{2}}\chi^{2}
\left({H}^{ \dagger  }H \right)-\lambda \left({H}^{ \dagger  }H \right)^{2}  \biggr) .
\end{split}\label{eq:sdedg}
\end{eqnarray}
Here, we redefine the scalar field as $\chi \rightarrow \phi=\chi \sqrt{12}$ and 
obtain the action for the Higgs sector as follows:
\begin{eqnarray}
\begin{split}
S_{{\rm gravity}+{\rm Higgs}} \supset &\int { { d }^{ 4 }x\sqrt { -\overline{g} } }  \biggr( \frac { M_{\rm pl}^{2}}{ 2}\overline{R }
-\frac { 1 }{ 2 } \overline{ g }^{ \mu \nu  }{ \overline{\nabla}  }_{ \mu  }{  \phi}{ \overline{\nabla}   }_{ \nu  } \phi  \\ &
-\frac { 1 }{ 2 } \overline{ g }^{ \mu \nu  }
{ \overline{\nabla}    }_{ \mu  }{ H}^{ \dagger  } {\overline{\nabla} }_{ \nu  }H-\frac{\Lambda_{\rm b}}{144M_{\rm pl }^{4}}\phi^{4} 
-\frac{M^{2}_{\Phi }}{12M_{\rm pl }^{2}}\phi^{2}
\left({H}^{ \dagger  }H \right)-\lambda \left({H}^{ \dagger  }H \right)^{2}  \biggr) .
\end{split}\label{eq:sdedg}
\end{eqnarray}

The transformed Higgs potential can be given by 
\begin{equation}
V\left({H}^{ \dagger  }H  \right) =\lambda_{\Lambda}\phi^{4}+\lambda_{M}\phi^{2}
\left({H}^{ \dagger  }H \right)+\lambda \left({H}^{ \dagger  }H \right)^{2},
\end{equation}
where the Higgs mass parameter and the cosmological constant become marginal operators with zero scaling dimensions
and they are fixed at the UV cut-off scale as
\begin{equation}
\lambda_{\Lambda}\left(\mu_{\rm UV}\right)= \frac{\Lambda_{\rm b}}{144M_{\rm pl }^{4}},\quad
\lambda_{M}\left(\mu_{\rm UV}\right)=\frac{M^{2}_{\Phi }}{12M_{\rm pl }^{2}}.
\end{equation}
The renormalisation group (RG) runnings of $\lambda_{\Lambda}$, $\lambda_{M}$ and $\lambda$  are 
determined by the one-loop $\beta$-functions~\cite{Gonderinger:2009jp,EliasMiro:2012ay}
\begin{eqnarray}
{ \beta  }_{\lambda_{\Lambda}} &=&\frac { 1}{ \left( 4{ \pi  }\right)^{2} }\biggr[ 
 2{ \lambda  }_{ M }^{ 2 }+20{ \lambda  }_{ \Lambda }^{ 2 }  \biggr] , \\
 { \beta  }_{\lambda_{M}} &=&\frac { 1}{ \left( 4{ \pi  }\right)^{2} }\biggr[
\frac { 1 }{ 2 }  { \lambda  }_{ M }\left( 12{ y }_{ t }^{ 2 }-3{ g' }^{ 2 }-9{ g }^{ 2 } \right) 
+4{ \lambda  }_{ M }\left( 3\lambda +2{ \lambda  }_{ \Lambda } \right) +4 { \lambda  }_{ M }^{ 2 }  \biggr] , \\
{ \beta  }_{\lambda } &=&\frac { 1}{ \left( 4{ \pi  }\right)^{2} }\biggr[
 \lambda \left( 12{ y }_{ t }^{ 2 }-3{ g' }^{ 2 }-9{ g }^{ 2 } \right) -6{ y }_{ t }^{ 4 }
+\frac { 3 }{ 4 } { g }^{ 4 }+\frac { 3 }{ 8 }{ \left( { g' }^{ 2 }+{ g }^{ 2 } \right)  }^{ 2 }
+24{ \lambda  }^{ 2 }+{ \lambda  }_{ M }^{ 2 }  \biggr]   .
\end{eqnarray}
where the one-loop running of the gauge coupling or 
the Yukawa coupling are not changed by 
the conformal transformation because the actions of the gauge sectors or the fermion sectors
are only rescaled via the conformal transformations.
The transformed Higgs potential has classical conformal symmetry and 
naturally realize the electroweak scale via the RG running behavior, i.e. the radiative symmetry breaking
where these theories are free from the electroweak hierarchy problem 
(see the reference~\cite{Iso:2009ss,Iso:2009nw,
Iso:2012jn,Hashimoto:2013hta} as the more detailed discussion).
Now, we found out that the classically conformal theory~\cite{Bardeen:1995kv} 
which solves the electroweak hierarchy problem is 
closely related with the modified gravity action of Eq.~(\ref{eq:sdedg}).

Following Polyakov's arguments~\cite{Polyakov:1982ug} 
it is found that the RG running effects of the dilaton or graviton would suppress the 
Higgs boson mass or cosmological constant in Eq.~(\ref{eq:sdedg}).
Although the equivalence between the quantum theories in the Einstein and Jordan 
frames is still under debate, $f(R)$ gravity theories in the Einstein and Jordan frame are equivalent on shell at the quantum level.
The (off shell) quantum corrections are ambiguous, but
the equivalence of the 
effective potential or renormalization group equations have been shown (see, e.g. the reference~\cite{Ruf:2017xon,Ohta:2017trn}).
Thus, the conformal transformation is also effective for 
the RG running effects and the modified gravity would relax the electroweak hierarchy.

\section{Conclusion}
\label{sec:conclusion}
In the present paper, we have discussed the gravitational relaxation scenarios for the electroweak hierarchy problem. 
We have clearly shown that the conformal transformation has an essential role to understand 
what modified gravity theory solves the hierarchy problem.
If the electroweak scale is naturally realized by dynamical and running behavior of 
dilatonic scalar field or scaling parameter, 
the modified gravity theory 
can relax the electroweak hierarchy problem.
The running gravitational relaxation models are closely related with 
the classically conformal theory and are 
theoretically attractive as suggested by Polyakov's arguments~\cite{Polyakov:1982ug}.
Also, the vacuum energy sequestering scenario can be recognized as the 
dilatonic suppression.
The modified gravity theory has the possibility of solving the electroweak hierarchy problem,
but require specific constraints on the gravity action with dilatonic scalar fields 
and it might provide non-trivial effects including metric instability and ghost 
on the observed Universe.

\bibliographystyle{JHEP}
\bibliography{relaxation}

\end{document}